\documentclass[prl,amssymb,floatfix,superscriptaddress,nofootinbib]{revtex4}
\usepackage{amsmath,amssymb,color}
\usepackage{graphicx}
\usepackage{slashed}
\usepackage{graphicx}
\usepackage[caption=false]{subfig}

\usepackage{leftidx}
\usepackage{cancel}
\usepackage[normalem]{ulem}
\usepackage[shadow,obeyDraft]{todonotes}

\newcommand{\Veff}{V_{\text{eff}}}
\newcommand{\leff}{\lambda_{\text{eff}}}
\newcommand{\ie}{i.e.}
\newcommand{\eg}{e.g.}
\newcommand{\MP}{M_p}
\newcommand{\NN}{\mathcal{N}}
\newcommand{\Hcal}{\mathcal{H}}

\newcommand{\be}{\begin{eqnarray}}
\newcommand{\ee}{\end{eqnarray}}

\def\ben{\begin{equation}}
\def\een{\end{equation}}
\def\bena{\begin{eqnarray}}
\def\eena{\end{eqnarray}}

\begin{document}
\begin{flushright}
MPP-2015-191\\
ICCUB-15-019
\end{flushright}
\title{Electroweak vacuum stability and inflation via non-minimal derivative couplings to gravity}

\author{Stefano Di Vita}
\email{divita@mpp.mpg.de}
\affiliation{Max-Planck-Institut f\"ur Physik, F\"ohringer Ring 6, 80805 M\"unchen, Germany}

\author{Cristiano Germani}
\email{germani@icc.ub.edu}
\affiliation{Institut de Ci\`encies del Cosmos (ICCUB), Universitat de Barcelona, Mart\'i Franqu\`es 1, E08028 Barcelona, Spain}

\begin{abstract}
  We show that the Standard Model vacuum can be stabilized if all particle
  propagators are non-minimally coupled to gravity.
  This is due to a Higgs-background dependent redefinition of the
  Standard Model fields: in terms of canonical variables and in the large Higgs field
  limit, the quantum fluctuations of the redefined fields are suppressed
  by the Higgs background. Thus, in this regime, quantum corrections
  to the tree-level electroweak potential are negligible. Finally,
  we show that in this framework the Higgs boson can be responsible
  for inflation. Due to a numerical coincidence that originates from
  the CMB data, inflation can happen if the Higgs boson mass,
  the top mass, and the QCD coupling lie in a region of the parameter space
  approximately equivalent than the one allowing for electroweak vacuum stability in the Standard Model.
  We find some (small) regions in the Standard Model parameter space in which the new interaction ``rescues'' the electroweak
  vacuum, which would not be stable in the Standard Model.
\end{abstract}

\maketitle
\section{Introduction}

The discovery of a light Higgs boson~\cite{discovery} represented the last step
towards the complete knowledge of the parameters of the Standard Model (SM) of particle
physics.
The state-of-the-art vacuum stability analyses of~\cite{divita, buttazzo} (see also references therein) showed that
the experimentally measured SM parameters are such that the SM picture can be
consistently extrapolated all the way up to the Planck scale $\MP=2.435\times10^{18}{\rm GeV}$
(\ie~where the effect of gravity can no be longer be neglected).
However, if no new degrees of freedom are advocated up to the Planck scale
and the effect of trans-Planckian physics is considered negligible, absolute stability of SM vacuum
is disfavored \cite{buttazzo}: a global minimum
is likely to be developed at large field values, rendering the electroweak (EW) vacuum metastable,
even though its lifetime is longer than the age of the universe.\footnote{
  Additional effects due to the expansion of the universe could enhance the EW vacuum decay
  probability, see~\cite{spencer}.} 
One may wonder whether the ultimate fate of the EW vacuum is rescued by some stabilization
mechanism that might come into play at high (or even trans-Planckian) energy scales.
On the other hand, as discussed in \cite{toni_last}, the EW vacuum metastability could
be the essential ingredient in order to avoid the quantum instability of the DeSitter solution
that our Universe seems to approach.
 
A very intriguing coincidence would be that the SM effective potential stays
positive during inflation, so that reheating can be accommodated within the SM~\cite{riottogiudice}.
An even more interesting scenario is that SM Higgs boson itself acts as the inflaton.
This can \eg~achieved by conformally coupling the Higgs boson to gravity as in the
so-called ``Higgs Inflation" of~\cite{fedorloop}. However, in this case, new (\ie~non-SM and
non-gravitational) degrees of freedom inevitably participate to inflation, or at
least, to the transition from inflation to the EW vacuum~\cite{alexriotto} (see
e.g.~\cite{giudice} for a completion of the Higgs inflation of~\cite{fedorloop}).

If instead the Higgs boson is kinetically coupled to curvature, as in the ``new Higgs inflation''
of~\cite{new}, no new degrees of freedom are necessary in the inflationary regime \cite{nico}.
In this paper we will consider an extension of such scenario and analyze the effect
of quantum corrections.
Specifically, we will introduce derivative interactions of the curvature
tensors to the kinetic terms of the SM fields, uniquely chosen in such a
way that no new degrees of freedom are introduced.
The feature of these new gravitational interactions is to change the
normalizations of the SM fields in a way that depends on the value of the
background Higgs field and on a new parameter.
At ``small'' background Higgs field values the theory effectively behaves
just like the ordinary SM. On the contrary, at ``large'' field values
the normalization becomes non-negligible, leading to an approximate decoupling
of the Higgs boson fluctuations.
This will be the key ingredient in order to $i)$ stabilize the SM effective
potential $ii)$ allow the Higgs boson to inflate the primordial Universe.

We will see however that, given the current cosmological and particle data, the choice of parameters that generate a successful inflation almost coincide with the choice of parameters that would anyway stabilize the EW vacuum in the SM.

\section{Quantum Analysis}
\subsection{Higgs-gravity system}
The model we are going to consider extends the one proposed in~\cite{new} in the context
of inflation, where the Higgs-gravity sector is chosen to be (we use the ``mostly plus'' signature)
\be\label{theory0}
\mathcal{L}=\int d^4x \sqrt{-\bar g} \,\Big[\frac{1}{2}M_p^2 \bar R - \left(\bar g^{\mu\nu}-\frac{ \bar G^{\mu\nu}}{M^2}\right){\cal D}_\mu \Hcal^\dag  \mathcal{D}_\nu \Hcal
 - V(\Hcal^\dag \Hcal)
\Big]\,
\ee
where $\bar R$ and $\bar G^{\mu\nu}$ are respectively the Ricci scalar (gravity
in vacuum is not modified) and the Einstein tensor,  $\Hcal$~denotes the complex Higgs doublet,
${\cal D}_\mu$ is the standard covariant derivative (under spacetime and $SU(2)\times U(1)_Y$),
and the potential is 
$V(\Hcal^\dag \Hcal) \simeq \lambda (\Hcal^\dag \Hcal)^2$ 
(up to a cosmological constant term and the quadratic term, which is negligible in the region we are
interested in).

The canonical momentum conjugated to the Higgs doublet is (on a spacelike hypersurface)
\be\label{recan}
\pi_{\cal H}\equiv \frac{\delta {\cal L}}{\delta\dot{\cal H}}=-2\sqrt{-g}\left(g^{\mu\nu}-\frac{ G^{\mu\nu}}{M^2}\right)n_\mu\partial_\nu{\cal H}\ ,
\ee
where $n_\mu$ is an arbitrary timelike unit four-vector with zero vorticity. Because in this system the
Higgs boson is non-canonical, imposing the standard equal time commutation rules one has~\cite{wilc}
\be\label{commu}
\left[{\cal H}({\bf x})^\dag,\dot{\cal H}({\bf y})\right]=\frac{1}{2}i\hbar\, \frac{\delta^{(3)}({\bf x-y})}{\NN}\ ,
\ee
where
\be\label{NN}
\NN\equiv-\sqrt{-g}\left(g^{tt}-\frac{G^{tt}}{M^2}\right)
\ee
in some coordinates adapted to $n_\mu$.
While it might be possible to work with the non-canonical Higgs boson with
commutations rules \eqref{commu} (see e.g.~\cite{wilc} for the Higgs inflation case of \cite{fedorloop}) we will instead
work with canonical fields following the approach of~\cite{fed2}.

In order to canonicalise the Higgs boson (and later on also the fermions and vectors of the theory) we use the following result:
defining a new metric
\be
g_{\alpha\beta}=\bar g_{\alpha\beta}+\epsilon_{\alpha\beta}\ ,
\ee
for small $\epsilon_{\alpha\beta}$ we have the perturbative expansion
\begin{align}
\label{EHvariation}
\int d^4x \sqrt{-g} R & = \int d^4x \sqrt{-\bar g}\bar R+\frac{\delta [\sqrt{-g}R]}{\delta g_{\alpha\beta}}\Big|_{g=\bar g}\epsilon_{\alpha\beta}+{\cal O}(\epsilon^2) \nonumber\\
& =\int d^4x \sqrt{-\bar g}\bar R-\bar G^{\alpha\beta}\epsilon_{\alpha\beta}+{\cal O}(\epsilon^2)\ ,
\end{align}
where the second term in the last equality comes from the standard variation of the Einstein-Hilbert action.
We thus see that, if we choose the disformal metric~\cite{sloth,romanian}
\be\label{disformal}
g_{\alpha\beta}=\bar g_{\alpha\beta}-\frac{{\cal D}_\alpha{\cal H}^{\dag}{\cal D}_\beta{\cal H}}{M^2 M_p^2}\ ,
\ee
and truncate at first order in the covariant derivatives, the non-minimal derivative coupling in \eqref{theory0}
cancels against the second term in \eqref{EHvariation}. 
However, the potential term ``remembers'' the original lagrangian, as we shall show.

The determinant of the metric is expanded in an analogous manner as
\be\label{7}
\sqrt{-g}=\sqrt{-\bar g}\left(1+\epsilon^\alpha{}_\alpha+{\cal O}(\epsilon^2)\right)\ ,
\ee
where indices are contracted with $\bar g^{\alpha\beta}$. Plugging in our choice for $\epsilon_{\alpha\beta}$ and inverting in favour of $\sqrt{-\bar g}$  
we then have
\be
V({\cal H}{\cal H}^\dag)\sqrt{- \bar g} = V({\cal H}{\cal H}^\dag)\sqrt{- g}\left(1+\frac{{\cal D}_\alpha{\cal H}{\cal D}^{\alpha}{\cal H}^\dag}{M^2 M_p^2}+\text{higher-covariant-derivatives\ interactions}\right)\ .
\ee

Summarizing, in terms of the disformal metric~\eqref{disformal} and at first order in the covariant derivatives,
the lagrangian~\eqref{theory0} reads
\be\label{theory}
\!\!\!\!\!{\cal L}\simeq\int d^4x \sqrt{-g}\Big[\frac{1}{2}M_p^2 R-\left(1+\frac{({\cal H}^\dag{\cal H})^2}{4 \Lambda^4}\right){\cal D}_\mu{\cal H}^\dag{\cal D}^\mu{\cal H}
-V({\cal H}^\dag{\cal H})\Big]\ ,
\ee
where $\Lambda\equiv \Lambda_t \lambda^{-1/4}$ at the classical level and $\Lambda_t= \sqrt{M M_p}$.
In \eqref{theory}, non-renormalisable interactions of the vector fields with the Higgs boson and 
higher-derivative interactions are neglected while all the self-interactions of the Higgs are kept, including the non-renormalisable ones (up to two-derivatives)
As we will discuss later on, the non-renormalisable self-interactions of the Higgs boson, after canonical normalisation in a
non-trivial Higgs background field, will be truncated at the renormalisable level. This is
consistent with our approximation of neglecting all non-renormalisable interactions that are
suppressed by a large Higgs boson background (as we shall consider). Finally, note that the
disformal transformation~\eqref{disformal}, when applied to the other SM
fields, will, again, only introduce higher-(covariant)-derivative interactions.

The theory~\eqref{theory}, seems to lose tree-level perturbative
unitarity when the potential reaches the transition value
$\Lambda_t^4$~\cite{nico}. However, perturbative unitarity is actually not lost. Indeed, at the same scale
a non-negligible gravitational background is generated, leading to a kinetic mixing
between the graviton and the Higgs boson~\cite{nico}.  Upon diagonalization of the
Higgs-graviton system, one discovers that the unitarity violation scale is actually
background dependent. Specifically, for a background in the zero-momentum limit
but for large field values (corresponding to large occupation number), one finds that the scale of perturbative
unitarity violation rises from $\Lambda_t$ to 
$\sim M_p$ during inflation~\cite{nico, sloth, yuki}.\footnote{Note that this background
  coincides with a DeSitter spacetime which is approximately a Friedman-Robertson-Walker
  inflating spacetime.}$^,$\footnote{While preparing this paper the Authors
  in~\cite{yuki_last}, by studying 2 by 2 scatterings but considering \emph{only} cubic
  interactions, found that the perturbative unitarity violating scale might be below
  $M_p$ but still well above the inflationary scales. However, as the Authors themselves
  admit, this result cannot be trusted until the quartic vertex are also included in the analysis.}
In the spirit of effective field theory one could also include in the diagonalized
system all possible higher-dimensional operators suppressed by the background
dependent cutoff that are compatible with its symmetries. However, as we shall be
only interested in background Higgs field values always far below the cutoff, we will consistently
neglect all of them.\footnote{
  The assumption here is that there is a UV complete theory with a non-trivial vacuum of which
  \eqref{theory0} is the low energy effective field theory. In addition, the
  theory \eqref{theory0} may be non-Wilsonian and self-unitarize \cite{uvself}. In this
  case there are no extra operators to be added, unless generated by loops.}

The effect of \eqref{disformal} is to ``integrate out"
the background transverse graviton by the use of the tree-level Einstein equations.
Neglecting Planck-scale suppressed longitudinal graviton fluctuations we are only left with a source generated by
a large number of background transverse gravitons (Coulomb-type field
strength), while transverse graviton fluctuations are gauged away by
diffeomorphisms~\cite{nico}, just as it would be for the electromagnetic field
coupled to a source. 
Specifically, in the zero-momentum limit (neglecting all the other SM fields),
and by using the classical Einstein equations
\be
\bar G_{\mu\nu}= \frac{T_{\mu\nu}}{M_p^2} \xrightarrow[p\rightarrow 0]{} -\frac{V}{ M_p^2}g_{\mu\nu}\,,
\ee
where all momenta are collectively denoted by ``$p$'', we have
\be
\NN\xrightarrow[p\rightarrow 0]{}1+\frac{V}{\Lambda_t^4}\,.
\ee
Far below the scale $\Lambda$ (small background field) $\NN\simeq 1$
and the Higgs-gravity system is well approximated by the SM. Far above
$\Lambda$ (large background field) $\NN\simeq \frac{({\cal H}^\dag{\cal H})^2}{4 \Lambda^4}$.
In the latter case one has to consider field redefinitions in order to to make
the commutator \eqref{commu} canonical and to be able to calculate
quantum corrections to the system in the usual way.

\subsection{Gauge-Fermions-Gravity sector}

Here we will extend the original model of \cite{new} by democratically coupling to gravity all
the SM kinetic terms \footnote{It
  is also interesting to point out that a non-minimally coupled axion to gravity
  can account for the missing Dark Matter, even for high inflationary energy
  scales without producing dangerous isocurvature perturbations \cite{javi}.} and by using a common suppression scale.
As for the fermions, the only non-minimal kinetic interaction that
does not introduce new degrees of freedom is again the term appearing
in~\eqref{theory0}~\cite{localization}.
Thus, we choose the coupling of the SM fermions (collectively called $\psi$) to be\footnote{Note that if fermions
  are supersymmetric partners of a non-minimally kinetically coupled scalar, they
  must have the coupling \eqref{ferm} \cite{fotis}.} 
\be\label{ferm}
{\cal L}_{\rm kin}^{\psi}=-\left(g^{\alpha\beta}-\frac{G^{\alpha\beta}}{M^2}\right)\bar\psi\gamma_\alpha{\cal D}_\beta\psi\ .
\ee

Analogously, there is only a non-minimal kinetic interaction to gravity for the gauge fields
that does not introduce new degrees of freedom (see e.g.~\cite{localization} and
references therein):
\be\label{gauge}
{\cal L}_{\rm kin}^{A}=-\frac{1}{4}\left(g^{\alpha\mu}g^{\beta\nu}+\frac{{\cal H}^\dag{\cal H}}{\Lambda_t^2}\frac{\ {}^{**}R^{\mu\nu\alpha\beta}}{M^2}\right){\rm Tr}F_{\alpha\beta}F_{\mu\nu}\ ,
\ee
where we collectively called $A$ the gauge vectors, $F$ denotes their field strengths, and
$^{**}R^{\mu\nu\alpha\beta}$ is the double-dual Riemann tensor.
Actually, the above interaction was shown in~\cite{nico} to be necessary in order
to avoid trans-Planckian gauge vector masses during inflation.\footnote{Note that
  in \cite{nico} the scale suppressing the Higgs
  boson was $\Lambda_M$. Here we prefer to use instead $\Lambda_t$ to have the
  transition to the non-minimally coupled system at the same point for all fields.}

We will now follow the discussion of \cite{fed2} and use the formalism of the
non-linear realization of symmetry breaking.
With a slight abuse of notation, we now parametrize ${\cal H}=\frac{h}{\sqrt{2}} {\cal U}$, where
${\cal U}=\exp\left[i\pi^a \tau^a\right]$ and $\pi^a$ are the
non-canonical Goldstone bosons. Similarly to the previous case, in the
zero-momentum limit, we have that 
\be
\left(g^{\alpha\mu}g^{\beta\nu}+\frac{{\cal H}^\dag{\cal H}}{\Lambda_t^2}\frac{\ {}^{**}R^{\mu\nu\alpha\beta}}{M^2}\right){\rm Tr}F_{\alpha\beta}F_{\mu\nu}
\xrightarrow[p\rightarrow 0]{} \left(1+\frac{h^2 V}{\Lambda_t^6}\right){\rm Tr}F^2\equiv \NN_A {\rm Tr}F^2\ .
\ee
As before we will use the approximation $\NN_A\simeq 1$ for $V \ll \Lambda_t^4$ and
$\NN_A\simeq \frac{h^2 V}{\Lambda_t^6}$ for  $V \gg \Lambda_t^4$. Thus, at small
background field values, the full system is approximately the SM.
In the next section we will consider the large field limit of this system.

\subsection{The large Higgs-background limit}

The canonically normalized Higgs boson $\chi$ is
\be
\chi=\int \sqrt{\NN} dh\ ,
\ee
which at large field values ($V\gg \Lambda_t^4$) is approximated by
\be\label{chiappr}
\chi\simeq\frac{\sqrt{\lambda}}{6}\frac{h^3}{\Lambda_t^2}\ .
\ee
The canonical Goldstone bosons will then be $\pi_{can}^a\simeq3\chi\pi^a$. 
The canonically normalized fermions $\psi_{can}$ and vectors $A_{can}$ will instead be
\be
\psi_{can}&\simeq&\frac{\sqrt{\lambda}h^2}{2\Lambda_t^2}\psi\cr
A_{can}&\simeq&\frac{\sqrt{\lambda}h^3}{2\Lambda_t^3}A\ .
\ee
In terms of these fields it is straightforward to derive an approximation
of our lagrangian in the large field limit. We work in the chiral representation
for the Higgs field (see a similar discussion in~\cite{fed2}) and neglect all the
higher-derivative operators and the operators suppressed by the
inverse power of the Higgs background. We obtain (for simplicity we drop the subscript ``$_{can}$''
unless otherwise specified):
\be
{\cal L}_{chiral}=-\frac{1}{2}(\partial\chi)^2-\frac{1}{g^2}H_1-\frac{1}{g'^2}H_2-L_{W/Z}+L_Y-U(\chi)\ ,
\ee
where
\be\label{SMhigh}
H_1&=&\frac{1}{2}{\rm Tr} W_{\mu\nu}^2\ ,\ H_2=\frac{1}{4}B_{\mu\nu}^2\cr
L_{W/Z}&=&\frac{\Lambda_t^2}{4}{\rm Tr} V_\mu^2\ ,\ L_Y=-\bar\psi^{L,R}\slashed{D}\psi^{L,R}\ ,
\ee
and, still for large $h$,
\begin{align}
V_\mu & = i W_\mu-i{\cal U}B_\mu^Y {\cal U}^\dag\ , \nonumber \\
W_\mu & = 2 W_\mu^a\tau^a\ ,\ W_{\mu\nu}=2\partial_{[\mu} W_{\nu]}+i[W_\mu,W_\nu]\ , \nonumber \\
B_\mu^Y & = B_\mu T^3\ ,\ B_{\mu\nu}=2\partial_{[\mu} B_{\nu]}\ .
\end{align}
It might seem puzzling to see no Yukawa interaction in \eqref{SMhigh}. However, in the
high energy limit, the quarks decouple from the Higgs. In fact, thanks to the
canonical normalization of the Goldstone bosons and the quarks, the Yukawa coupling
is suppressed by the large Higgs field: e.g.\ once the normalization of the fermions
is taken into account the Yukawa coupling reads $y_Q \frac{2\sqrt{2}\Lambda^4}{ h^2}\bar Q_L {\cal U}Q_R$.
A similar argument shows that no kinetic term for the Goldstone bosons enters in $V_\mu$. Therefore for large
$h$, the quarks decouple from the Higgs, as well as the gauge vectors, as it is clear from
\eqref{SMhigh}. In other words, the Higgs boson is decoupled from the other fields.
Conversely to the small field limit, where the masses of the $W/Z$ bosons are proportional
to the background, here their masses saturate at $\Lambda_t$.

The tree-level  Higgs potential in terms of $\chi$, at large field values, is simply
\be\label{U}
U(\chi) \equiv V(h(\chi)) = \lambda \frac{h(\chi)^4}{4} \simeq (m^2\chi)^{4/3}\ .
\ee
where $m = (9/2)^{1/4}\ \Lambda_t\lambda^{1/8}=(9/2)^{1/4}\ \Lambda \lambda^{3/8}$.

To calculate the one loop effective Coleman-Weinberg potential~\cite{coleman}, we need to know
the (field dependent) mass of $\chi$. It is a trivial computation to see that, for large
Higgs background field, $m_\chi^2=\frac{d^2U}{d\chi^2}\propto \frac{\Lambda^2}{h^2}$
and thus, under our approximations, will be taken to vanish. In addition, expanding the
potential \eqref{U} around the background $\chi_0$, \ie~$\chi=\chi_0+\delta\chi$ it is clear
that the only non-vanishing term is a tadpole and therefore all beta functions associated to
the self-Higgs interactions are (approximately) trivial. Thus, the effective potential
above the scale $\Lambda$ will be well approximated by its tree-level form and the scale
$m$ will not (approximately) run. More precisely, loop effects will be suppressed by the
large Higgs boson background.

\subsection{Matching and EW vacuum stabilization}

Far below the scale $\Lambda$ (which we refer to as ``region {\rm I}''), we can neglect
gravity and approximate the whole system with the SM. In this regime, we can calculate
the effective potential with the standard techniques, although in a gauge \emph{dependent}
way (for a recent discussion see e.g.~\cite{mihali} and references therein). Far from
the EW vacuum, the SM effective potential can be recast in the form
$\Veff(h) = \leff(h) h^4/4$, where $\leff(h)$ is the
\emph{effective quartic coupling} and its two-loop expression can be found in~\cite{buttazzo}.
Instead, far above $\Lambda$ (``region {\rm II}''), the canonical Higgs boson is approximately
decoupled and thus we can ignore any gauge-dependence. In this region, as we have just
showed, the effective potential is well approximated by its tree-level form,
parametrized by a background {\it independent} value of $m$. 

Assuming for a moment a sudden transition between the regions {\rm I} and {\rm II}
at $h_*=\sqrt{2}\Lambda$ (corresponding to $\chi_*\simeq 0.47 \Lambda$), the net effect is that
the gauge dependent $\leff(h)$ sharply converges
to the (gauge-independent) running coupling $\lambda(\sqrt{2}\Lambda)$ (we have implicitly
made the usual choice $\mu=h$ for the renormalization scale and used the fact
that $\Lambda$ depends weakly on the background Higgs value, as we shall see shortly).
 
Obviously, the transition between the two regions is not sharp. However, since
the $\NN$ factor changes with $h^4$, it is reasonable to assume that the
width of the transition region is ${\cal O}(1)$ GeV. Above roughly $10^5$ GeV the change
of the Higgs quartic coupling $\lambda(h)$ is very mild in a generic range
$\Delta h\sim {\cal O}(1)\ \rm GeV$~\cite{buttazzo}. Thus we expect the same to
happen in the transition region and this will be our working assumption.
In other words, we expect that a sharp transition between region {\rm I} and
{\rm II} is not going to be a bad approximation. 

The question still to be answered is whether the scale $\Lambda$ varies with the
background field value. For $\frac{({\cal H}^\dag{\cal H})^2}{4 \Lambda^4}\ll1$, the term
$\frac{({\cal H}^\dag{\cal H})^2}{4 \Lambda^4}{\cal D}_\mu{\cal H}^\dag{\cal D}^\mu{\cal H}$
can be considered as a self-interaction for the field ${\cal H}$. 
It is then an easy exercise to see that, in this regime, the scale $\Lambda$ runs very
weakly with the renormalization scale $\mu$: the only diagram (at one-loop)
generating the running of $\Lambda$ is the one involving the quartic Higgs
coupling and the running turn out to be $\frac{d\ln \Lambda}{d\ln\mu}=\frac{3}{4\pi^2}\lambda$.
As $\lambda$ is small and runs to even smaller values in the region we are interested in,
we can safely neglect the running of $\Lambda$ in our analysis.\footnote{Note that in absence
  of a potential term the scale $M$ would still enter via the higher-derivative operators (that
  we neglected here). Nevertheless, its running would be forbidden by a non-renormalization
  theorem~\cite{nonren}.}
A similar analysis also reveals that the scales appearing in~\eqref{ferm} and~\eqref{gauge}
run weakly.

In the small background field regime ($h\ll \Lambda$), the first non-SM
interaction in terms of $\cal H$ is of quartic-Galileon type \cite{gal}. Schematically,
this is $\frac{1}{\Lambda_M^6}{\partial\cal H}^\dag\partial{\cal H}\partial^2{\cal H}^\dag\partial^2{\cal H}$,
where $\Lambda_M=(M^2 M_P)^{1/3}$ is the scale at which perturbative unitarity is
violated at large momenta. Note that, at high momentum (but still at small field values), the Higgs boson is approximately
invariant under Galilean transformations. We assume here that the UV completion of the
theory at high momentum is still invariant under this approximate symmetry, therefore
it must involve only derivative operators that would not spoil the low-momentum analysis.\footnote{
  As an alternative approach, one could remove all interactions suppressed by $\Lambda_M$ by
  subtracting a covariant Galileon component (see \cite{covgal} for the definition)
  in the original lagrangian \eqref{theory0} so that the scale $\Lambda_M$
  is removed from the theory.}
Because of null results in the search of non-SM phenomena in collider
experiments at large momenta, we will constrain $\Lambda_M$ to be
above $\mathcal{O}(1)$ TeV. This implies $\Lambda \gtrsim  10^7\ {\rm GeV}$.

The scale $m$ we are interested in is then finally
\be\label{mdef}
m = (9/2)^{1/4}\ \Lambda \lambda_*^{3/8}
\ee
where the value of the running $\lambda$ at the transition point,
$\lambda_*\equiv\lambda(h_*)$, can be calculated at three-loop accuracy
following~\cite{divita, buttazzo, lambda}.

The running of $\lambda$ is mainly affected by the strong and Yukawa interactions.
Therefore, in our analysis we keep as free parameters (within a few standard deviations
from the current average values) the Higgs boson pole mass ($m_h$), the QCD coupling $\alpha_s$
evaluated at the $Z$ boson mass, and the top quark pole mass ($m_t$). The latest world average values are:
$\alpha_s=0.1185\pm 0.0006$ , $m_h=(125.09\pm 0.24){\rm GeV}$ and 
$m_t=(173.34\pm 0.76){\rm GeV}$ \cite{world}. Notice that the top quark pole mass suffers from an irreducible non-perturbative uncertainty of the order of~$\pm\Lambda_{\rm QCD} \simeq \pm 0.3$ GeV (see \eg~\cite{divita,buttazzo}). Furthermore, the relation between the top quark mass that is reconstructed at hadronic colliders, using Monte Carlo simulations, with its pole mass involves further subtleties, see \eg~\cite{hoang,moch} and references therein.
For simplicity, we approximate the top quark pole mass of \cite{moch}, corresponding to the experimental world average, with $m_t=(173.39\pm 1.05){\rm GeV}$ . To summarize, in this paper we use the following values  
\be\label{smvalues}
\alpha_s&=&0.1185\pm 0.0006\ ,\cr
m_h&=&(125.09\pm 0.24){\rm GeV}\ ,\cr
m_t&=&(173.39\pm 1.05){\rm GeV}\ .
\ee
As already mentioned in the introduction, with these values for the input parameters,
SM vacuum stability is disfavored:
the SM effective potential develops a global minimum at large
field values and the EW vacuum turns out to be metastable. This can be avoided in
our framework, provided the transition happens before the scale $h_0$ at which the
running coupling vanishes $(\lambda(h_0)=0$). Together with our lower bound on $\Lambda$,
the EW vacuum is stabilized if
\be\label{range}
10^7 \,{\rm GeV} \lesssim \Lambda = \frac{h_*}{\sqrt{2}}\ll \frac{h_0}{\sqrt{2}}\ .
\ee
For the central values we find $h_0\simeq 6\times 10^9\ {\rm GeV}$ and therefore
a value for $\Lambda$ can be accommodated such that the above equation is
satisfied and stability recovered.

\section{The Higgs boson as inflaton}  

Thanks to the gravitational enhanced friction mechanism \cite{new, romanian, yuki},
far above the scale $\Lambda$, if $\lambda_*$ is positive, the Higgs boson rolls very
slow down its own potential generating an almost DeSitter phase (inflation). This
happens because the non-minimal coupling of the Higgs boson's kinetic term to the
Einstein tensor increases the general relativistic kinetic energy loss (Hubble
friction) of the Higgs boson to gravity. 

The cosmic microwave background radiation (CMB), very precisely observed by the ESA
Planck satellite experiments \cite{planck}, is described by  the amplitude of the power spectrum \cite{yuki, pert1}
\be {\cal P}\simeq \frac{H^2}{8\pi^2\epsilon\ M_p^2}\simeq 2\times 10^{-9}\ ,
\ee 
the spectral index 
\be n_s=1-5\epsilon\,,
\ee 
and the tensor to scalar ratio 
\be
r=16\epsilon\ .
\ee 
In the high friction regime in which $V(h(\chi))\gg \Lambda_t^4$ (precisely the regime in
which quantum corrections are under control) we have \cite{yuki} (see \cite{nico} for
full non-approximate formulas)
\be
\epsilon=\frac{8}{3}\frac{M^2}{H^2}\frac{M_p^2}{h_I^2}\ ,
\ee 
where the Hubble constant is $H^2=\frac{V}{3 M_p^2}$ and $h_I$ is the Higgs background
value during inflation.

During inflation the Universe expands $e^N$ times. In order to have a
successful inflation, inflation should last between $50$ to $60$ e-foldings. The relation
of the number of e-foldings ($N$) with the slow-roll parameter $\epsilon$ is \cite{nico}
\be 
N= \frac{1}{3}\left(\frac{1}{\epsilon}-1\right)\ .
\ee
Once the number of e-foldings is fixed, $n_s$ and $r$ are uniquely determined.
For $N$ ranging from $50$ to $60$ we have
\be\label{ns}
n_s&=&\left\{
	\begin{array}{ll}
		0.966  & \mbox{if } N=50 \\
		0.972 & \mbox{if } N=60
	\end{array}
\right.\ ,\cr
r&=&\left\{
	\begin{array}{ll}
		0.106  & \mbox{if } N=50 \\
		0.088 & \mbox{if } N=60
	\end{array}
\right.\ ,
\ee
which are completely independent of $\lambda_*$.

The values in \eqref{ns} fit within one sigma the latest Planck data analysis \cite{planck}.
Note that, if we were not in the high friction limit we could have had higher values
for $r$, as shown in \cite{nico}.
\footnote{While replying to the Referee's comments, the new BICEP2/KECK analysis appeared in \cite{bicepn} claiming an upper bound for $r<0.07$ at 2-sigma level. We note that this can be achieved in our model for $N\simeq 75$ while still being within 2-sigma level from the central value of $n_s$ from Planck. However, as already discussed, since our analysis is weakly dependent in $N$, we will, for simplicity, only consider the value of $N$ compatible with the central value of $n_s$ obtained by Planck.}

Although, as we said, the cosmological parameters are independent of $\lambda_*$, this is
not true for the scale $M$ and the Higgs boson background value during inflation ($h_I$).
However, in the high friction limit, the constant $m$ entering the potential~\eqref{U}
is completely fixed by the CMB. It is easy
to find that
\be\label{hm0}
m&\simeq&\frac{5.38\times 10^{15}}{(1+N)^{5/8}}{\rm GeV}\ .
\ee
Similarly, the value of the canonically normalized Higgs field during inflation is
\be
\chi_I\simeq 3.96\times 10^{18}\ {\rm GeV}\sqrt{N+1}\ .
\ee
This is what we expect: in chaotic inflation the value of the canonical inflaton must be trans-Planckian.

A last condition we have to impose is that inflation happens above the transition scale, \ie~that
\be
\label{chiI}
\chi_I\gg \chi_*\ .
\ee
As we discussed, $\chi_* \simeq 0.47 \Lambda$. Therefore, \eqref{chiI} is satisfied
provided $\Lambda\ll M_p$, which is actually a consistency condition in quantum gravity~\cite{qg}.

\subsection{EW vacuum stability and inflation}
\begin{figure}
  \centering
  \subfloat{
    \includegraphics[width=.5\textwidth]{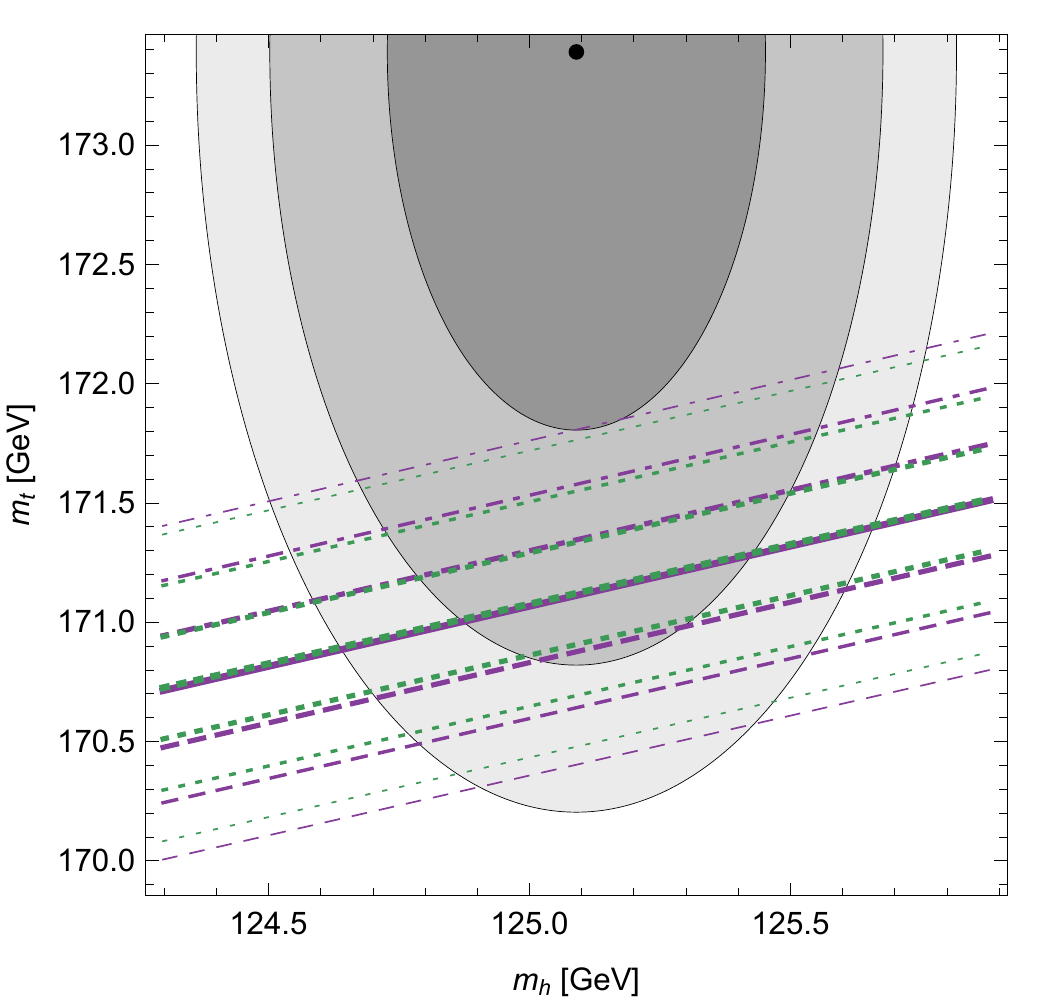}}
  \subfloat{
    \includegraphics[width=.5\textwidth]{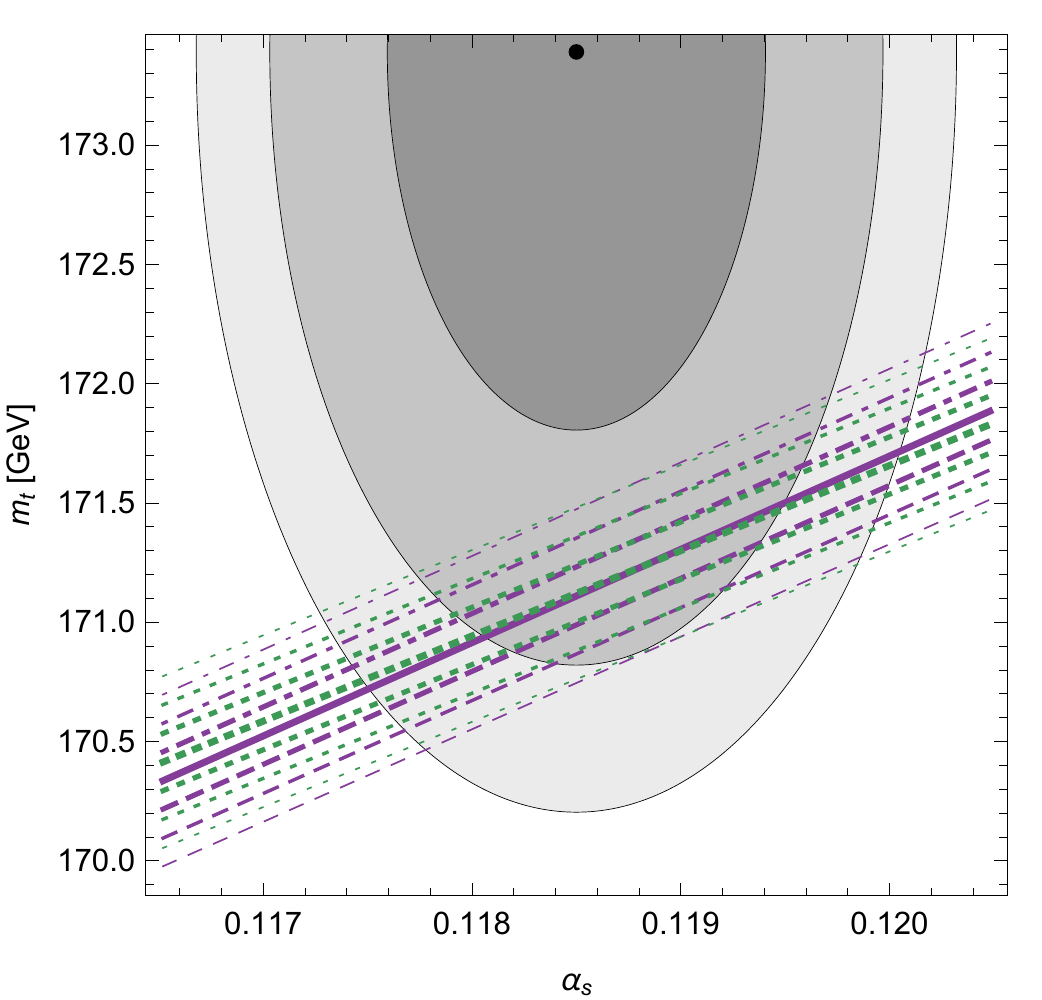}}
  \caption{
    The left (right) panel displays (in purple) the upper boundary of the allowed
    region in the $(m_h\,,m_t)$ ($(\alpha_s\,,m_h)$) plane for several choices of
    $\alpha_s$ ($m_h$). The thick solid lines correspond to the central value of
    the fixed parameter; the dot-dashed lines (with decreasing thickness from the
    thick line) represent choices that are larger than the central value by
    (respectively) $1,2,3\,\sigma$;
    the dotted lines represent choices that are smaller than the central value by
    (respectively) $1,2,3\,\sigma$.
    %
    The dotted green lines show the stability bound in the SM (taken from~\cite{buttazzo})
    corresponding to the closest purple curve.
    The gray ellipses are the $68\%\,,95\%$ and $99\%$ probability regions for the
    parameters on the axes. See the text for more details.
}
  \label{plots}
\end{figure}

If we restrict ourselves to the range $50\leq N\leq 60$, our equations are only weakly dependent
on $N$. Therefore, within our working precision, we can safely fix $N=51$, which corresponds
to the central value of the spectral index observed by Planck \cite{planck}, $n_s=0.968$.
From \eqref{hm0} we have
\be\label{hm}
m&\simeq&4.55\times 10^{14}\ {\rm GeV}\ .
\ee
Note that in the SM $\lambda$ is small at the EW scale, decreases quite rapidly but then varies very slowly
with the running scale.
Since $m = (9/2)^{1/4} \Lambda \lambda_*^{3/8}$, one expects, when the conditions for inflation
and EW vacuum stabilization are met, $\Lambda_t$ not to differ from $m$ (and $h_*$) by more
than 1-2 orders of magnitude (and indeed we checked that this is the case).

We have performed scans of the experimentally allowed region in the $(m_h\,, m_t \,,\alpha_s)$
parameter space (see~\eqref{smvalues}) in order to assess whether it is possible to achieve simultaneously \emph{i)}
successful inflation and \emph{ii)} EW vacuum stabilization.
In each of our scans we have fixed one of the three parameters
to the central value of its latest determination and we have varied the remaining
two within the corresponding $3\,\sigma$ regions, \ie~the most interesting from
a phenomenological point of view.
For each point we have checked whether or not the condition~\eqref{hm} can be satisfied for some $\Lambda$
allowed by the constraint~\eqref{range}. We have repeated the same scan considering $\pm 1,2,3\,\sigma$ variations of the parameter that we fix.

In Fig.~\ref{plots} we show our results for the scans in which either $\alpha_s$ or $m_h$ are kept fixed.
The left panels displays (in purple) the upper boundary of the allowed region in the $(m_h\,,m_t)$ plane for
several choices of $\alpha_s$. The thick solid line corresponds to the central value $\alpha_s=0.1185$.
The dot-dashed lines represent choices that are larger than the
central value by $1,2,3\,\sigma$ (in order of decreasing thickness from the solid line). 
Analogously, the dotted lines are the results
we obtain with $\alpha_s$ smaller by $1,2,3\,\sigma$ than the central value.
We also show, for convenience, the $68\%\,,95\%$ and $99\%$ probability regions for
the parameters on the axes, assuming them to be independent gaussian variables with mean
and standard deviation as given in~\eqref{smvalues}.
Close to each line, the stability bound in the SM, from~\cite{buttazzo}, is displayed in dotted
green lines for the same value of the fixed parameter.
%
Analogous comments apply to the right panel, in which we show the boundary lines in the $(\alpha_s\,,m_h)$ plane
obtained after having fixed the Higgs boson mass.

From both plots we see that the condition for successful inflation (and EW vacuum stability) is relatively
close to that for absolute stability in the SM.
This can be understood as a consequence of our requirements from cosmology, that eventually
fix the numerical value of $m$ to $\mathcal{O}(10^{14})$ GeV, and of the beta-functions of the SM.
For parameters that favor stability of the SM vacuum, the running of the Higgs quartic
coupling in the SM is such that $\lambda$ and $\leff$ do not considerably differ.
The requirement (for inflation) that $\lambda$ is positive (and large enough)
in order to allow \eqref{mdef} and \eqref{hm} to be satisfied, turns out to be, due to the SM running,
essentially equivalent to requiring $\lambda$ (and therefore $\leff$) to be always positive.
To be more precise, one sees from the left panel that the two bounds are
basically parallel in the $(m_h,m_t)$ plane and that a slight crossing
happens as $\alpha_s$ is varied (see also the right panel). For the central value
the two bounds overlap; for smaller $\alpha_s$ the region that allows inflation
is contained in the one that allows SM vacuum stability; for larger $\alpha_s$
the non-canonical kinetic interaction ``rescues'' the EW vacuum, which would
be metastable in the SM, and allows for successful inflation from the Higgs sector.

We do not show the analogous plot of the scan in the $(m_h,\alpha_s)$ plane, but
we briefly comment on the results. Consistently with what one would expect,
having a somehow light top quark is necessary in order to satisfy our constraints.
What we find is that unless we choose $m_t$ at least roughly 1.5 standard
deviations \emph{below} its average value, it is not possible to have inflation
and vacuum stabilization.
On the other hand, for a light top quark the criteria are mildly dependent
on $m_h$ and $\alpha_s$: for $m_t=171.29$ GeV most of the upper-right corner
is allowed, while for for $m_t=170.24$ GeV the allowed region covers essentially
the whole parameter space we analyzed.

As final remark, the viability of our scenario strongly relies on
the top quark \emph{pole} mass being smaller than the current world average. While
the Higgs boson pole mass is measured with a remarkable precision at the LHC and can be
approximately considered as a given parameter, a direct precise measurement of the
top quark pole mass suffers from considerable theoretical uncertainties and
this fact still provides some room for speculations.
One could alternatively use the $\overline{MS}$ top mass as an input parameter
for the analysis, bypassing in principle all the issues with the pole mass,
but the precision of its current experimental determination is not such that
any conclusive statement can be made (see \eg~\cite{divita,alekhin} for a
discussion in the context of the SM vacuum stability bound).

\section{Conclusions}

If the SM has a non-minimal kinetic coupling to gravity in a way that no
new degrees of freedom are added, we showed that the EW vacuum can be
stabilized even for the central values of the SM parameters (which, within the
sole SM would imply a metastable EW vacuum).

In this scenario, the Higgs boson can be considered as responsible for cosmic
inflation (as already shown at the classical level in~\cite{new}).
We showed that, within two standard deviations from the current
central values of the most relevant SM parameters ($m_h,m_t$ and $\alpha_s$), there
exist points such that \emph{i)} the
EW vacuum is stabilized due to an approximate decoupling of the Higgs field at
large background field values and \emph{ii)} inflation is achieved in
compatibility with current data~\cite{planck}.%

In general, the allowed parameter space turns out to be essentially similar
to the one that allows for SM vacuum stability due to a numerical coincidence,
\ie~the fact that cosmological data set the scale at which the new interaction
becomes relevant to a large value of $\mathcal{O}(10^{14-16})$.
More precisely we observe that for given $\alpha_s$ the corresponding boundary
lines are parallel in the $(m_h,m_t)$ plane, while variations of $\alpha_s$
generate a slight crossing: as the QCD coupling increases with respect to
its current central value, the region we find becomes somehow larger than the
SM vacuum stability region, signaling that the new interactions ``rescues''
the EW vacuum.

\section*{Acknowledgments} 
C.G. was supported by the Ramon y Cajal program and MDM-2014-0369 of ICCUB (Unidad de Excelencia 'Mar\'ia de
Maeztu'). C.G. wishes to thank Sarah Folkerts for discussions on loops with derivative couplings, Kyle Allison for correspondence, Alex Spencer-Smith for discussions on the the effective potential during inflation, Gia Dvali and Alex Kehagias for comments on the first draft of the paper. C.G. wishes to thank Humboldt Foundation and Ludwig-Maximilians-Universit\"at for support during the initial part of the project. S.D.V.\ and C.G.\ wish to thank Dario Buttazzo and Alessio Notari for discussions on the gauge dependence of the effective potential.

\end{document}